# Towards a Cognitive Handoff for the Future Internet:

A Holistic Vision

Francisco A. González-Horta, Rogerio A. Enríquez-Caldera, Juan M. Ramírez-Cortés, Jorge Martínez-Carballido
Department of Electronics, INAOE
Tonantzintla, Puebla, México
{fglez, rogerio, jmram, jmc}@inaoep.mx

Eldamira Buenfil-Alpuche
Faculty of Engineering
Polytechnic University of Guerrero State (UPEG)
Taxco, Guerrero, México
e-mail: eldamira@gmail.com

*Abstract—* Current handoffs are not designed to achieve multiple desirable features simultaneously. This weakness has resulted in handoff schemes that are seamless but not adaptive, or adaptive but not secure, or secure but not autonomous, or autonomous but not correct, etc. To face this limitation, we initiated a research project to develop a new kind of handoff system which attains multiple purposes simultaneously by using context information from the external and internal handoff environment. We envision a cognitive handoff as a multipurpose, multi-criteria, environment-aware, and policy-based handoff that trades-off multiple objectives to reach its intended goals. This paper presents a conceptual (soft) model of cognitive handoffs using a holistic approach. We applied the proposed model to identify cognitive handoff performance parameters and tradeoffs between conflicting objectives. We argue that cognitive handoffs are the archetype of handoffs for the future Internet.

## I. INTRODUCTION

A handoff is intended to preserve the user communications while different kinds of transitions occur in the network connection. Thus, a handoff is the process of transferring communications among radio channels, base stations, IP networks, service providers, mobile terminals, or any feasible combination of these elements.

Significant desirable handoff features mentioned in the literature are, e.g., seamless [1], adaptive [2], autonomous [3], secure [4], and correct [5]; however, many others can be found in the vast literature of handoffs: transparent, reliable, flexible, robust, balanced, immune, fast, soft, smooth, lossless, efficient, proactive, predictive, reactive, QoS-based, power-based, location-aided, time-adaptive, intelligent, generic, etc. Despite the rich variety of desirable handoff features, two important problems remain unsolved: (1) how can be combined different desirable features into a single handoff process so that it can achieve many purposes simultaneously? (2) how to define every desirable feature so that ambiguity and subjectivity can be reduced?

This gap in knowledge about handoffs has produced a number of single-purpose schemes that successfully achieve one attractive feature but completely ignore others; e.g., seamless handoffs with poorly or null adaptation to other scenarios or technologies [6]; adaptive handoffs that do not consider any security goal [2]; secure handoffs that ignore user autonomy [4]; etc. Also, there is a growing confusion in literature about similar features; e.g., accurate-correct, fast-timely, smooth-seamless, robust-reliable, etc. In order to reduce misuse and ambiguity of these attributes is convenient to associate a qualitative property (purpose) and quantitative measures (objectives and goals) to each desirable feature. By doing so, we can qualify and quantify their performance individually or in comparison with others. Major contributions of this research paper include:

*1) A new holistic vision of handoffs.* Many handoff solutions follow a reductionist approach; i.e., they achieve one desirable feature, use a small amount of handoff criteria, and work only in very specific scenarios. Although these simplistic solutions provide understanding and control of particular situations, we have seen how they quickly become special cases of more general models. Thus, we claim that the handoff problem for the future Internet requires holistic solutions, achieving multiple desirable features, using a diversity of context information, and adapting to any handoff scenario.

*2) A new conceptual model for cognitive handoffs.* We propose a new kind of handoff that is multipurpose, multi-criteria, context-aware, self-aware, policy-based, and trades-off multiple conflicting objectives to reach its intended goals. This paper provides the conceptual model and its first level of functional decomposition.

## II. SINGLE-PURPOSE VS MULTI-PURPOSE HANDOFFS

Dr. Nishith D. Tripathi in his outstanding thesis work published in 1997 [7] probably was the first author in considering a handoff that can simultaneously achieve many desirable features. His inspiring work served for many years as a basis for developing high performance handoffs; however, the complexities of handoff scenarios from 1997 to present days have changed significantly. For instance, the handoff concept changed from simple lower-layer transitions between base stations and channels to more elaborated cross-layer transitions among networks, providers, and terminals. The limited scope of Tripathi's handoff concept has brought in consequence that his algorithms and models become today special cases of more general models. Holism is relevant in this way to provide a long-term solution for the handoff problem. Another author who describes several desirable handoff features is Nasser et al. [8] in 2006. Both, Tripathy and Nasser, described various desirable features, but they did not make any difference among features, purposes, objectives, and goals. A handoff model needs a clear distinction to such former concepts.

The holistic vision to the handoff problem has also been studied by Dr. Mika Ylianttila in his exceptional thesis work [9] published in 2005. He presented a holistic system architecture based on issues involved in mobility management areas (e.g., mobility scenarios, handoff strategies, handoff control, handoff algorithms, handoff procedures, mobility protocols, mobility parameters, performance measures, and handoff metrics). The work of Ylianttila improved the architecture of handoff issues that Pahlavan [10] published in 2000. However, these architectures have some drawbacks: i) they did not include the context management problem in their models; ii) they did not mention the tradeoffs that handoffs should consider in a multi-objective scenario; and iii) their architectures are based on types of issues and not in the functionality aspects of the handoff process.

Besides the above related work, we use two criteria to classify handoff schemes that are approaching to cognitive handoffs: the number of desirable features they achieve and the amount of context information they use. Handoff schemes, like the ones proposed by So [6] and Zhang [11], achieve only one desirable feature using limited context information; they provide seamless handoffs between particular network technologies and specific mobility scenarios. The schemes proposed by Siddiqui [12] and Hasswa [13] use broad context information, but they are focused only in one feature (seamlessness).

Conversely, the solutions proposed by Sethom [4] and Tuladhar [14] provide seamless and secure handoffs on a variety of handoff scenarios because they use broad context. The schemes proposed by Singhrova [2] and Chen [15] achieve seamless and adaptive mobility, but they cannot adapt to any handoff scenario because they use limited context. Finally, the scheme proposed by Altaf [16] achieves seamless, secure, soft, and adaptive handoffs, but just between WiMAX and 3G networks because they use limited context.

Considering this tendency, it will be common to observe in the near future a new generation of handoffs that can achieve many desirable features using broad handoff context information. In current literature, none architecture, model, or algorithm is reported to have this property.

Regarding the related work of standardization bodies, like the IEEE 802.21 and the IETF MIPSHOP, we observed that they are focusing in seamless heterogeneous handoffs; they are not taking into account the vast diversity of desirable features that handoffs could have. The IEEE 802.21 workgroup has approved three task groups to face very particular handoff scenarios: the IEEE 802.21a for security extensions to media independent handovers, the IEEE 802.21b for handovers with downlink only technologies, and the IEEE 802.21c for optimized single radio handovers. We believe they are following a reductionist approach, but they lack the holistic vision of cognitive handoffs. Emmelman, in [17], discusses ongoing activities and scopes of these standardization bodies.

III. THE COGNITIVE HANDOFF HOLISTIC VISION

*A. Origin of Single-Purpose Handoffs*

The thoughtful study of handoffs started in the early 1990s with the first generation (1G) cellular networks (e.g. AMPS [18]). These networks provided seamless conversations while the mobile phone switched between channels and base stations. The decision to perform a handoff was made only on a signal strength basis, but the handoff execution should be imperceptible to users. For this reason, the AMPS system required that the handoff gap be no more than 100 ms to avoid the possibility of dropping a syllable of speech [18]. These traditional handoffs are single-purpose/single-criterion or seamless/signal strength.

*B. Major Challenges in the Future Internet*

*1) Multidimensional Heterogeneity:* A major trend in future communication systems is the coexistence of multiple dimensions of heterogeneity integrated into a seamless, universal, uniform, ubiquitous, and general-purpose network. This future Internet will be seamless if it hides heterogeneity to users, universal if it can be used by anyone with any terminal, uniform if it is an all-IP network, ubiquitous if it is available anywhere and anytime, and general-purpose if it can provide any service. We divide heterogeneity into five dimensions as illustrated in Fig. 1 and explained in the next paragraphs. The arrows going down from the service provider dimension to the user mobility dimension depict two different handoff scenarios created by instantiating objects in each dimension.

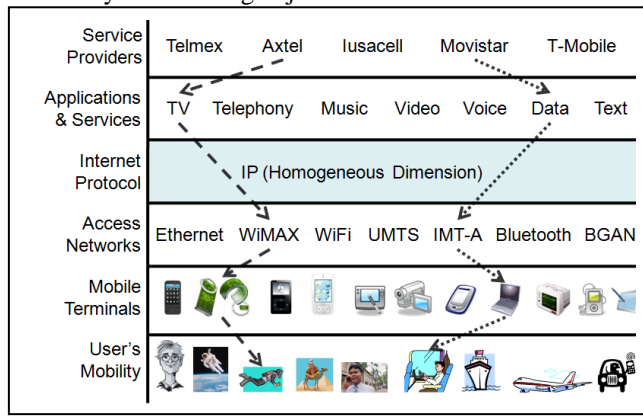

Figure 1. Multidimensional heterogeneity in the future Internet.

*a) Diversity on service providers and operators:* Offer different classes of services, billing models, security policies, and connection prices. They deploy different wireless technologies around the world and make roaming agreements and alliances with other providers and operators.

*b) Variety of applications and services:* Intend to fulfill the distinct ways of human communication; e.g., voice, video, data, images, text, music, TV, telephony, etc.

*c) Several access network technologies:* Include wired and wireless access technologies [19]; e.g., Ethernet, Bluetooth, WiMAX, WiFi, UMTS, MBWA, IMT-2000, GPRS, GSM, EDGE, LTE/SAE, DVB-HS, etc. They differ in terms of electrical properties, signaling, coding, frequencies, coverage, bandwidth, QoS guarantees, mobility management, media access methods, packet formats, etc.

*d) Plethora of mobile user terminals:* Users can be humans, machines, or sensors. Terminals for machines are integrated parts of machines. Sensor terminals collect information from networked sensors [19]. Terminals for humans are mobile and multimode, equipped with telecommunication capabilities and different saving energy characteristics; they change its factor form from those looked like computers (laptops, netbooks) to those looked like cell phones (PDAs, smartphones).

*e) Numerous user mobility states:* Network terminals can be located anywhere – in space, on the ground, under the ground, above water, underwater, and they can be fixed in a geographic position or moving at any speed – pedestrian, vehicular, ultrasonic [19].

Nowadays, no handoff solution exists which comprehensively addresses the entire scale of heterogeneity. Moreover, multidimensional heterogeneity has three main attributes: is inevitable, is the source of great amounts of context information, and produces an infinite number of handoff scenarios.

*2) Ubiquitous Connectivity:* It enables connectivity for anyone or anything, at any time, from anywhere. A myriad of wireless access technologies are spread across the entire world overlapping one another but avoiding interferences among them. Two requirements for ubiquitous connectivity are: (i) to develop scalable architectures to integrate any number of wireless systems from different service providers [20] and (ii) to develop smart multimode mobile terminals able to access any wireless technology [21].

*3) Cognitive Mobility:* It allows roaming mechanisms where the user is always connected to the best available network, with the smaller number of handoffs, service disruptions, user interventions, security threats, and the greater number of handoff scenarios.

### C. External and Internal Handoff Environment

We envision a cognitive handoff as a process that is both context-aware and self-aware. This implicates to make the handoff process aware of its external and internal environment. We borrowed the term 'cognitive' from Dr. Dixit vision of cognitive networking [22]. He defines *cognitive networking* as an intelligent communication system that is aware of its environment, both external and internal, and acts adaptively and autonomously to attain its intended goals. We believe cognitive handoffs not only should behave adaptively or autonomously to attain its intended goals, but also seamlessly, securely, and correctly.

On one hand, the external environment is directly related with all the external entities that provide a source of context information to the handoff process. These entities are users, terminals, applications, networks, and providers; a cognitive handoff should adapt to any kind of these entities. These entities maintain a strong cyclic relationship as follows: users interact with terminals, terminals run applications, applications exchange data through networks, networks are managed by providers, and providers subscribe users. The cyclic relationship of external entities suggests that all external context information emanates just from these five basic entities and no more; hence, if we ignore information of any of these entities, the handoff process will not adapt properly to all the scenarios. Therefore, a cognitive handoff should consider all the five entities.

On the other hand, the internal environment is another source of context and it is directly related with the behavior or performance of handoffs. This behavior directly depends on the desirable features of handoff. Next, we identified and describe five major desirable features which are considered highly significant for the current and future scenarios.

### D. Multiple Desirable Features of Handoff

*1) Seamlessness:* It means to preserve the user communications before, during, and after the handoff thus reducing service degradation or interruption. Service degradation may be due to a continuous reduction in link quality, network quality, handoff quality, QoS guarantees, and energy savings. Service interruption may be due to excessive degradations or a "break before make" approach.

*2) Autonomy:* This desirable feature is closely related to seamlessness. A handoff is autonomous, automatic, or autonomic when no user interventions are required during a handoff in progress. However, this does not mean that user interventions are not required in handoffs. It is good that users participate in the handoff configuration process by defining their preferences, priorities, or necessities; but, it is convenient that users can perform this activity offline to prevent any distraction during online communications.

*3) Security:* We say a handoff is secure if not new threats appear along the handoff process and security signaling traffic does not overload the network and degrades the communication services. This is a very challenging task, but if optimization techniques are used together with our model it could be shown that by minimizing handoff latency, authentication latency, and signaling overload, the risk of new threats appearance may be reduced.

*4) Correctness:* A handoff is correct if it keeps the user always connected to the best available network with the smaller number of handoffs; this is similar to the Gustaffson's vision of ABC defined in [23]. We consider that the best network is the one that is sufficiently better and consistently better. Furthermore, correctness can bring other additional features to the handoff process:

- *Beneficial*: if quality of communications, user expectations, or terminal power conditions get improved after handoff.
- *Timely*: if handoff is executed just in time; i.e., right after target is properly selected and before degradations or interruptions occur.
- *Selective*: if it properly chooses the best network among all the available networks.
- *Necessary*: if it is initiated because of one imperative or opportunist reason.

- *Efficient*: if it selects the most appropriate method, protocol, or handoff strategy, according to the types of: handoff in progress, user mobility, and application.

These handoff attributes derived from correctness, take special relevance during the decision-making phase, where it must be decided why, where, how, who, and when to trigger a handoff.

*5) Adaptability:* An adaptable handoff should be successful across any handoff scenario. A handoff is successful if it achieves a balance of every desirable feature at a minimum level of user satisfaction.

*E. Structure of Handoff Context Information*

The handoff context information is extensive, heterogeneous, distributed, and dynamic. It supports the whole operation of the handoff process and the achievement of multiple desirable features. Therefore, such context information should be arranged in a clear structure. Table I and Table II show the structure of handoff context information according to a pair of criteria: the source of context and the class of information respectively. The sources of context originated in the external handoff environment support context-awareness while the one originated in the internal environment (the handoff process itself) will provide self-awareness.

TABLE I. STRUCTURE FOR SOURCE OF CONTEXT INFORMATION

| |
|---|
| *User context:* This context allows users to customize the handoff according to their own needs, habits, and preferences. It includes: user preferences, user priorities, user profiles, user history, etc. |
| *Terminal context:* Allows the deployment of QoS-aware handoffs, power-based handoffs, and location-aided handoffs:<br>(a) *Link quality:* Received signal strength (RSS), signal to noise ratio (SNR), signal to interference ratio (SIR), signal to noise and interference ratio (SNIR), bit error rate (BER), block error rate (BLER), co-channel interference (CCI), carrier to interference rario (CIR), etc.<br>(b) *Power management:* Battery type (BT), battery load (BL), energy-consumption rate (ECR), transmit power in current (TPC), transmit power in target (TPT), power budget (PB), etc.<br>(c) *Geographic mobility:* Velocity (Vel), distance to a base station (Dist), location (Loc), direction (MDir), coverage area (GCA), etc. |
| *Application context:* This context includes the QoS requirements of active applications: Lost packets (LP), delayed packets (DP), corrupted packets (CP), duplicated packets (DuP), data transfer rate (DTR-goodput), packet jitter (PJ), out-of-order delivery (OOD), application type (AppT), etc. The consideration of these QoS parameters makes provisions for application-aware handoffs. |
| *Network context:* This context is needed to avoid selecting congested networks (befor handoff), to monitor service continuity (during handoff), and to assess the handoff success by measuring network conditions (after handoff): Network bandwidth (NBW), network load (NL), network delay (ND), network jitter (NJ), network throughput (NT), network maximum transmission unit (NMTU), etc. |
| *Provider context:* Connection fees, billing models, roaming agreements, coverage area maps, security management (AAA), types of services (data, voice, video), provider preferences, and provider priorities. A negotiation model may be required to equate the differences between service providers, network operators, and mobile users. |
| *Handoff performance context:* Call blocking (CB), call dropping (CD), handoff blocking (HOB), handoff rate (HOR), handoff latency (HOL), decision latency (DLat), execution latency (ExLat), evaluation latency (EvLat), handoff type (HOType), elapsed time since last handoff (ETSLH), interruptions rate (IR), interruption latency (IL), degradations rate (DR), degradation latency (DL), degradation intensity (DI), utility function (UF), signaling overload (SO), security signaling overload (SSO), improvement rate (ImpR), application improvement rate (AppImpR), user improvement rate (UsrImpR), terminal improvement rate (TermImpR), successful handoff rate (SHOR), imperative handoff rate (IHOR), opportunist handoff rate (OHOR), dwell time in the best (DTIB), authentication latency (AL), detected attacks rate (DAR), online user interventions rate (OUIR), tardy handoff rate (THOR), premature handoff rate (PHOR), etc.his context allows users to customize the handoff according to their own needs, habits, and preferences. It includes: user preferences, user priorities, user profiles, user history, etc. |

TABLE II. STRUCTURE FOR CLASS OF INFORMATION

| |
|---|
| *Handoff criteria:* Network discovery, decision-making, and performance evaluation. Some examples of handoff criteria include variables or parameters from the external/internal environment such as RSS, NL, BL, LP, HOL, Vel, connection price, etc. |
| *Handoff metrics:* Mathematical models used to measure several significant tasks of the handoff process; for instance, the quality of links, the quality of communications, the quality of different networks, the quality and quantity of handoffs, the quality of different providers, the achievement of user preferences, the power budget of a mobile terminal, the geographic mobility of a user, etc. Handoff metrics may combine a variety of handoff criteria and help any specific handoff algorithm to make optimal decisions. |
| *Performance measures:* Set of handoff metrics that are used to quantify performance of communications, performance of networks, performance of handoffs, and to evaluate the degree of achieving a handoff objective. |
| *Handoff policies:* Users and providers define a series of policies to the handoff operation. Policies define and specify rules for making handoff decisions in any particular situation; for instance, what to do if the link quality drops below a level required for an acceptable service. User and provider may have different views of the handoff process; provider may be interested in QoS while user in connection charges. Both points of view must be consistently integrated into a single handoff policy management database. |
| *Handoff constraints:* Conditions that must be satisfied in a particular handoff scenario and used to control the handoff operation by keeping performance parameters within specific limits. For instance, for a seamless handoff process, the delay has to be kept within certain boundaries; for real-time applications a delay of 50 ms could be acceptable, whereas non-real-time applications might accept delays as long as 3-10 sec [9]. |
| *Handoff configuration:* Defines preferences, priorities, and other configuration parameters required to customize the handoff operation. Typically, the configuration information is organized in a handoff profile linked to a particular user, provider, and terminal and should be initially performed offline either by the user, the provider, both or an auto-configuration setup. But, depending on the type of handoff algorithm, different configuration parameters may be required to be initialized, e.g. thresholds, timers, hysteresis, weights, etc. |

*F. Cognitive Handoff Conceptual Model*

Once we have established and justified the necessity for developing a new handoff system, we present our conceptual model based on the statement that "a cognitive handoff should intend to achieve multiple desirable features and be aware of its entire environment by using information coming from multiple context domains". Fig. 2 depicts this basic idea by interconnecting multiple desirable features with multiple context domains that we already explained separately in III.D and III.E.

The purpose of this model is to help people debate and discuss about the complexity of cognitive handoffs. Thus, topics of discussion would be related to level of complexity, correlation among desired features and context data, and the possibility of establishing handoffs as a multi-objective optimization problem as well as to give specifications for practical implementations. Used in this way this model is not intended for predicting, designing, or implementing cognitive handoffs, but for understanding and explaining such difficult and complex process All the above issues have not been addressed in the handoff literature; therefore , in effect, the purpose of this conceptual model is being achieved. Models like the one we present here are validated by credibility, and credibility comes from the way in which the cognitive maps are built and the clarity it represents most of the opinion's experts [24]. In the next section we provide some advances towards the development of cognitive handoffs.

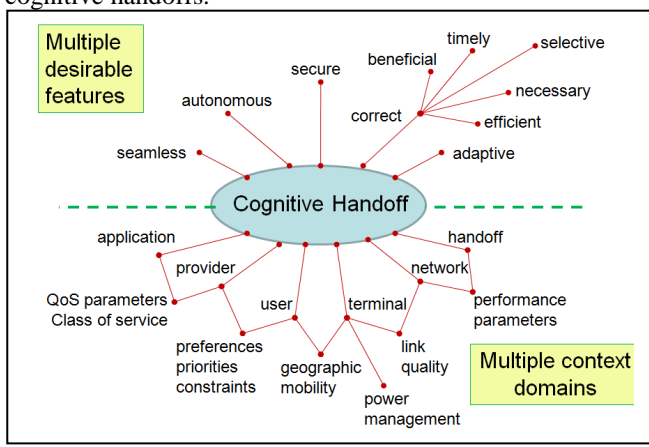

Figure 2.   Cognitive handoff conceptual model. The desired features to achieve determine the context data to use and vice versa.

## IV. COGNITIVE HANDOFF MODEL AT WORK

### A. Cognitive Handoff and Complex Systems

Cognitive handoffs are complex adaptive systems because: (1) they exhibit a complicated hierarchical structure (e.g., a power saving system is part of a network discovery system, which is part of a handoff system, which is part of a mobility system, which is part of a wireless communication system, and so on, but also a power saving system is part of the decision system, which is part of the handoff system, and so on); (2) the whole cognitive handoff system achieves purposes that are not purposes of the parts (e.g., a cognitive handoff purpose is to maintain the continuity of services, but this purpose is not defined in any of the parts or subsystems of the cognitive handoff system); and, (3) the handoff environment is dynamic and therefore adaptability is a desired handoff feature.

### B. Correlating Desired Features and Context Data

With respect on whether all previously described context data are necessary to describe limitations on the model; one has to realize that the usage of certain context parameters depends on the desirable features being implemented and the context data available in a moment will allow to accomplish or not a particular desired feature. Thus, we need to state a correct relationship or dependence between each desirable handoff feature and the subset of context data necessary to be accomplished. We made a correlation between desired features and context data by transforming desired features into purposes, purposes into objectives, objectives into goals, and goals into context data. For the sake of space in this manuscript the mentioned correlation will be shown in our next paper subtitled "Model-driven methodology and taxonomy of scenarios" already accepted for publication [25].

### C. Advances for a Practical Implementation

The cognitive handoff system, represented in Fig. 2 by the oval in the middle, can be expanded into several sub-systems by using a functional decomposition approach [26]. Fig. 3 shows the main functional sub-systems for cognitive handoffs represented in ovals: handoff control algorithm, network discovery, handoff decisions, handoff execution, handoff evaluation, and handoff context information management. We briefly describe them:

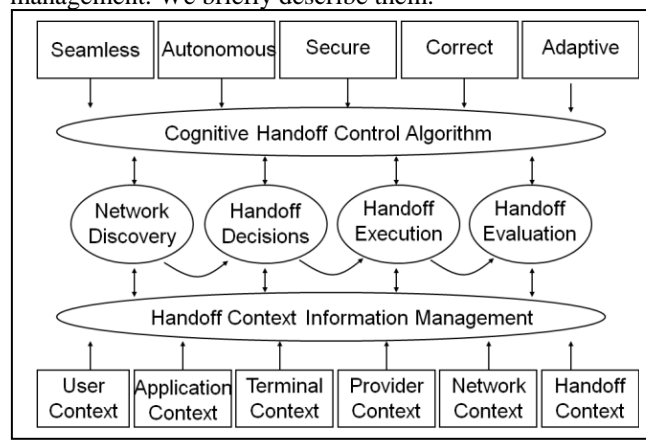

Figure 3.    Functional decomposition model. The desired features provide purposes, objectives, and goals to achieve, while context domains provide the information needed to attain such goals.

- *Handoff Control Algorithm*: This is the main director of the handoff procedure. The entity which implements the control algorithm is called Handoff Control Entity (HCE). There should be one HCE in every user terminal and also there may be many others distributed across the network infrastructure. HCEs are agents that cooperate and compete to take a particular handoff to succeed.
- *Network Discovery*: This is the system for detecting and discovering available access networks. An available network is a reachable and authorized network considered for an eventual handoff.
- *Handoff Decisions*: The handoff decisions system is intended to answer the questions of why, when, where, how, and who should trigger the handoff. Typically this system has focused only in where and when to handoff [27]. The holistic vision extends the scope of handoff decisions.
- *Handoff Execution*: This system is intended to change the physical and logical connection from one network to

another, from one provider to another, or from one terminal to another. This change requires the most effective method, protocol, or strategy according to the current handoff scenario. The MIPSHOP group at IETF and the IEEE 802.21 standard are creating tools for implementing media-independent handoffs since 2003.

- *Handoff Evaluation*: This system measures the achievement of every desirable handoff feature and decides whether the executed handoff was successful or not. The evaluation results should be delivered after the handoff execution but within strict time constraints, thus this task is proactively distributed along the handoff process.
- *Handoff Context Information Management*: This system is intended to collect the distributed handoff context data, transform the data in information, and redistribute this information to the HCEs which are responsible for making handoff decisions and control.

Discovery, decisions, execution, and evaluation systems can be viewed as sequential stages of the handoff process; however, the context manager is a background process which permanently supplies the handoff control entities with fresh information about the handoff environment.

### D. Cognitive Handoff Performance Measures

The performance evaluation of cognitive handoffs requires a performance metric for each handoff purpose and a graphical representation to visualize multivariate data [28]. These metrics combine mathematically several performance measures that are associated to every handoff purpose. It is possible that metrics can normalize heterogeneous data into a single value representing the performance of each handoff purpose. Moreover, metrics can also be designed as utility functions so that greater values are better and all values are on the same scale. Fig. 4 exemplifies a radar graph comparing the performance of multiple handoff purposes simultaneously. We say that if all measures are within a boundary circle of acceptable quality, then the cognitive handoff is successful, otherwise the handoff is defective and outliers should be corrected.

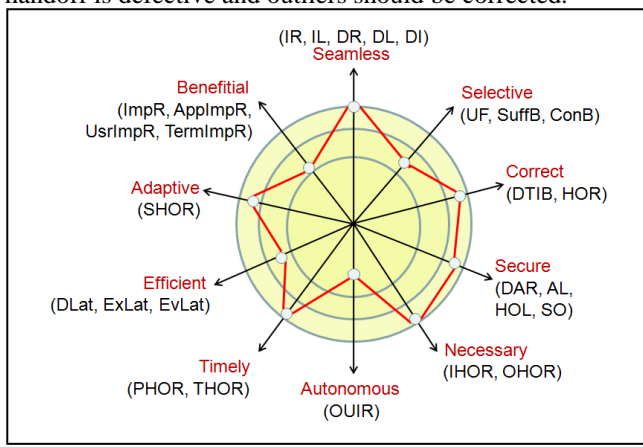

Figure 4. Functional decomposition model. The desired features provide purposes, objectives, and goals to achieve, while context domains provide the information needed to attain such goals.

### E. Formulating the Cognitive Handoff as a MOP

Let F be the set of desirable handoff features and C be the set of context data. We say that a context variable $V_i \in C$ is *correlated* with a desired feature $f \in F$ if and only if a change on the value of $V_i$ impacts on the purpose of f. For instance, some changes on the value of SNR may degrade or improve the link quality and impact on the purpose of seamlessness that is to maintain the continuity of services; thus, we say that SNR is correlated with seamlessness.

Let $V_f$ be the set of correlated variables with f, where $V_i \in V_f \subseteq C$. We say that $V_i$ is *positively correlated* with f if and only if increments on the value of $V_i$ produce improvements on the purpose of f and decrements on $V_i$ produce degradations on the purpose of f. For instance, increments on SNR improve the link quality, which improves the service continuity of seamlessness, and conversely, decrements on SNR degrade the link quality, which degrades the service continuity of seamlessness. Therefore, SNR is positively correlated with seamlessness.

↑SNR → ↑LINKQUALITY → ↑SEAMLESSNESS
↓SNR → ↓LINKQUALITY → ↓SEAMLESSNESS

We say that $V_i$ is *negatively correlated* with f if and only if increments on the value of $V_i$ produce degradations on the purpose of f and decrements on $V_i$ produce improvements on the purpose of f. For example, increments on BER degrade the link quality, which degrades the service continuity of seamlessness, and conversely, decrements on BER improve the link quality, which improves the service continuity of seamlessness. Therefore, BER is negatively correlated with seamlessness.

↑BER → ↓LINKQUALITY → ↓SEAMLESSNESS
↓BER → ↑LINKQUALITY → ↑SEAMLESSNESS

The set $V_f$ is partitioned in two subsets $V_f^+$ and $V_f^-$ where $V_f^+$ is the set of variables positively correlated with f and $V_f^-$ is the set of variables negatively correlated with f.

Every $V_i$ has a weight $W_i$ associated to its priority where $W_i \in \Re[0,1]$ and $\sum W_i = 1$. Let **v** represent the vector of variables $\mathbf{v} = (V_1, V_2, \ldots, V_m)$, then the *objective function* for the desired handoff feature f is defined by

$$f(\mathbf{v}) = \sum (K+W_i)\log(V_i^+) - \sum (K+W_i)\log(V_i^-) \quad (1)$$

where K is a scaling factor so that small changes on the context variables reflect big changes on $f(\mathbf{v})$.

In this general objective function, $V_i^+$ and $V_i^-$ are positively and negatively correlated variables of f. The objective function $f(\mathbf{v}) : \Re^m \to \Re$ is a utility function that we want to maximize because in desirable features the higher the value the best.

Considering k different objective functions $f_i$ that we want to maximize simultaneously where some of them may be in conflict, then the multi-objective optimization problem (MOP) can be stated as the problem of

$$\text{Maximize } \{f_1(\mathbf{v}), f_2(\mathbf{v}), \ldots, f_k(\mathbf{v})\} \quad (2)$$
$$\text{Subject to } \mathbf{v}_l \leq \mathbf{v} \leq \mathbf{v}_u,$$

where $\mathbf{v}_l$ and $\mathbf{v}_u$ represent the vectors of lower and upper values of the tolerance range for each variable.

*F. Tradeoffs between Conflicting Objectives*

A cognitive handoff is designed to achieve multiple purposes, objectives, and goals simultaneously. In the space of handoff objectives, we can distinguish between those with complementary nature and those with competitive nature. Complementary objectives can be simultaneously optimized without any conflict between them, but competing objectives cannot be simultaneously optimized, unless we find compromised solutions, largely known as the tradeoff surface, Pareto-optimal solutions, or non-dominated solutions [29]. We describe several tradeoffs to consider in a multi-objective handoff scheme:

a) *(Max. DTIB and Min. HOR)*: There is a tradeoff between maximizing the time to stay always best connected (DTIB) and minimizing the number of handoffs (HOR). The conflict arises because in a dynamic environment the best network is changing frequently and stochastically; thus, to maximize DTIB is necessary to make frequent handoffs as soon as a new best is available. This increase in the number of handoffs creates a conflict with minimizing HOR.

b) *(Min. DLat and Max. SHOR)*: This tradeoff is between minimizing the handoff decisions latency (DLat) and maximizing the number of successful handoffs (SHOR). The conflict emerges because the less time elapsed to make decisions will necessary lead to reduce the number of successful handoffs. For example, in case of imperative handoffs, DLat is reduced but this may lead to select an incorrect target because the selection time is also reduced.

c) *(Max. Sizeof-ContextInfo and Min SO)*: This is a tradeoff between minimizing the handoff signaling overload (SO) and maximizing the amount of handoff context information to be managed by the handoff control entities. The conflict arises because broad handoff information is required to attain multiple desirable features, but this will increase the amount of signaling traffic in the network.

d) *(User and Provider Preferences)*: Several conflicts may appear due to differences between provider and user preferences. For instance, providers may prefer networks within its own administrative domain while users may prefer networks with lower charges even if they are owned by other service providers; users may prefer a Mobile Controlled Handoff (MCHO) while providers may prefer Network Controlled Handoffs (NCHO). Conflicts like these require a balance between different interests. Handoff protocols like Mobile Assisted Handoff (MAHO) and Network Assisted Handoff (NAHO) try to balance the handoff control [7].

V. CONCLUSION

Handoffs are an integral component of any mobile-wireless network from past, present, and future. Handoffs are transitions that change the data flows from one entity to another, where these entities may be radio channels, base stations, IP networks, service providers, and user terminals. The handoff process should exhibit several desirable features beyond seamlessness and should consider more context information beyond the signal strength. This is a common requirement to face the handoff scenarios of the future Internet.

The existing handoff schemes are not able to achieve a variety of attractive features and managing arbitrary amounts of context information. Therefore, we proposed a conceptual model to create handoffs of this kind. We characterized a cognitive handoff to be multipurpose, multi-criteria, context-aware, self-aware, and policy-based.

We claimed that our cognitive handoff model is holistic because it considers all the transition entities that may be involved in handoffs, all the external and internal sources of context, and considers many significant desirable features.

Using a functional decomposition approach, we divided the functional behavior of a cognitive handoff into six general modules: control algorithm, network discovery, handoff decisions, handoff execution, handoff evaluation, and context management. Each module has assigned a purpose to every feature and decomposed each purpose into objectives and goals. We applied the cognitive handoff model to define its performance parameters and significant tradeoffs between conflicting objectives.

As a future work, we are preparing another manuscript for presenting a new taxonomy of handoff scenarios and the model-driven methodology that we are using to develop cognitive handoffs. There is still much work to do before we can see cognitive handoffs practically implemented. The cognitive handoff project follows theoretical and practical avenues. A theoretical challenge is to further develop the cognitive handoff MOP to study the structure of the variables in the handoff context (e.g., continuous/discrete, deterministic/stochastic, etc.) and the types of constraints required to create a convex optimization problem. In the practical and Applicability Avenue, we have deployed temporal and geometric simulation models to observe and predict the behavior of cognitive handoffs with two conflictive objective functions; however, further development is required to demonstrate the feasibility and applicability of cognitive handoffs in complex scenarios.

ACKNOWLEDGMENT

F.A. González-Horta is a PhD candidate at INAOE and thanks the financial support received from CONACYT Mexico through the doctoral scholarship 58024.